\documentclass[11pt]{article}

\makeatletter
\def\ps@top{\let\@mkboth\@gobbletwo
     \def\@oddhead{\rm\hfil\thepage\hfil}\def\@oddfoot{}
     \def\@evenhead{}\let\@evenfoot\@oddfoot}
\def\@bibsetup{\itemindent=-\leftmargin}
\def\@citesep{; }
\def\@cite#1#2{({#1\if@tempswa , #2\fi})}
\def\@biblabel#1{\hfill}
\def\thebibliography#1{\section*{References\markboth
 {REFERENCES}{REFERENCES}}\list
 {[\arabic{enumi}]}{\settowidth\labelwidth{[#1]}\leftmargin\labelwidth
 \advance\leftmargin\labelsep
 \usecounter{enumi}\@bibsetup}
 \def\newblock{\hskip .11em plus .33em minus -.07em}
 \sloppy
 \sfcode`\.=1000\relax}
\renewcommand{\section}{\@startsection {section}{1}{\z@}{-3.5ex plus -1ex minus 
    -.2ex}{2.3ex plus .2ex}{\centering\large\bf}}
\renewcommand{\subsection}{\@startsection{subsection}{2}{\z@}{-3.25ex plus
    -1ex minus -.2ex}{1.5ex plus .2ex}{\centering\bf}}
\makeatother
\raggedright
\begin{document}
\vspace*{1in}
\begin{center}
{\large\bfseries Observations of O~($^1$S) and O~($^1$D) in Spectra of C/1999\,S4 (LINEAR)} \\ [25pt]
Anita L. Cochran and William D. Cochran \\
McDonald Observatory \\
University of Texas at Austin \\ [15pt]
Accepted for Publication in {\itshape Icarus} \\ [2in]
\end{center}

\begin{center}
ABSTRACT
\end{center}

We report on high spectral resolution observations of comet C/1999\,S4 (LINEAR)
obtained at McDonald Observatory in June and July 2000. 
We report unequivocal detections
of the O~($^1$S) and O~($^1$D) metastable lines in emission in the
cometary spectrum.  These lines are well separated from any telluric or
cometary emission features.  We have derived the ratio of the two red
doublet lines and show they are consistent with the predictions
of the branching ratio.  We also derived a ratio of $0.06\pm0.01$ for
the green line flux to the sum of the red line fluxes.
This ratio is consistent with H$_{2}$O as the dominant parent for
atomic oxygen.  We have measured the widths of the lines and show
that the widths imply that there must be some parent of atomic
oxygen in addition to the H$_{2}$O.

\vspace*{0.5in}

\newpage
\section{Introduction}

Oxygen was one of the most common elements in the solar nebula and therefore
oxygen should be a major constituent of cometary ices.
Much of the cometary oxygen is incorporated into H$_{2}$O ice with
additional oxygen in CO, CO$_2$, H$_{2}$CO, HCOOH, CH$_3$OH, etc.
As a comet is heated during its approach to the Sun, the sublimated
parent gases undergo
two-body and photodissociation reactions, producing the myriad of radicals
and atoms observed in the cometary spectrum.  One of the species
commonly detected is atomic oxygen. 

Figure~\ref{transit} shows an energy level diagram for atomic oxygen. 
The strong triplet at $\sim$1304\AA\ has been observed from above the
Earth's atmosphere with facilities such as IUE and HST (e.g. Weaver
{\it et al.} 1981). 
This resonance fluorescence line is useful for determining the total amount
of atomic oxygen in the
coma, but gives no clue to the O parent species since it requires merely that
O atoms in the ground 2p$^4$~$^3$P state be excited to the 3s~$^3$S$^0$
state by solar photons.

In the optical region of the spectrum, three important atomic oxygen transitions
exist.  These are the forbidden oxygen red doublet at 6300.304 and 6363.776\AA\ 
($^1$D -- $^3$P) and the green line at 5577.339\AA\ ($^1$S -- $^1$D).
Festou and Feldman (1981) argued effectively 
against solar resonance fluorescence or
dissociative recombination of CO$^+_2$ as excitation processes for the
green and red lines.
Instead, these transitions must arise from atoms produced directly in the
excited
$^1$S or $^1$D states by photodissociation of a parent molecule (i.e. these
lines represent ``prompt" emission).
In addition, the lifetimes of these states are short. 
The lifetime at 1\,{\sc au}
of the $^1$D state is about 150\,sec, while the $^1$S state lifetime is
less than 1 sec. 
As a result, O atoms produced in the $^1$S or $^1$D levels will decay quickly
to the ground state before the atom has had a chance to travel very far from
the location at which it was produced.
Thus, the red and green lines represent detailed
tracers of their parent's outflow.

Photodissociation of an oxygen-bearing parent species can produce oxygen
atoms in the ground $^3$P state or in the excited $^1$S or $^1$D states,
depending on the parent molecule and the nature of the solar photodissociation.
If the green line ($^1$S -- $^1$D) is observed then, provided that collisional
deexcitition is negligible, the red doublet must also
be present since every such transition produces an oxygen atom in the
$^1$D level and the red doublet is then the only decay pathway
available.  For O atoms
in the $^1$S level, 95\% decay via the green line and then the red doublet
and 5\% decay in the near UV 2977 and 2958\AA\ doublet.
However, the red doublet can exist without the green line since oxygen atoms
can be created directly in the $^1$D state and decay to the $^3$P state.

While studies of the green and red transitions offer much valuable information
about the comae of comets, this information has not been utilized
much in the past because of difficulty in observing these lines.
All three lines are also present in telluric spectra and so observations
of comets must be obtained at high spectral resolving power to separate
the telluric and cometary lines (Magee-Sauer {\it et al.} 1988;
Combi and McCrosky 1991; Schultz {\it et al.} 1992; Schultz {\it et al.}
1993).   
Alternatively, some attempts have been
made to model the distribution of the telluric lines (mostly the
red doublet) and remove their
signal from the cometary lines (Spinrad 1982; Delsemme and Combi 1979, 1983).
In addition to contamination from the telluric lines, the 6300\AA\ red line
is also near the Q branch of one of the NH$_{2}$ bands.
However, moderately high spectral resolution is sufficient for
differentiating the oxygen and the NH$_{2}$.

Unambiguous detection of the green line is much harder than the red doublet. 
In addition
to the coincidence with the telluric green line, the cometary green
line is coincident with the cometary C$_{2}$ (1,2) P-branch.
The C$_{2}$ band is generally very strong and dense, making unambigous detection
of the green oxygen line very difficult.  High spectral and/or spatial
resolution has been used to detect the green line in comets with
varying degrees of success (W. Cochran 1984; Smith and Schempp 1989;
Morrison {\it et al.} 1997).  

Figure~\ref{examples} shows observations we have 
obtained of the spectral region around the green line for other comets.
These data are of high spectral resolving power.  A typical
spectrum for comet deVico is shown in the top panel.  The strong C$_{2}$ (1,2)
band head is marked.  All of the rest of the lines are attributable
to C$_{2}$.  The cometary and telluric O~($^1$S) lines are
resolved from one another in this spectrum, but the contamination
from C$_{2}$ is substantial.

In the bottom panel we show observations of comet Hyakutake.
These were obtained with still higher spectral resolving power.  In
addition, since the comet was very close to the Earth at the time
of the observations, the slit covered only the inner coma.  Since
C$_{2}$ has a longer scale length than O ($^1$S), the cometary oxygen line
is much more cleanly defined than for deVico.  However, it is still
apparent that there must be some amount of C$_{2}$ contamination.

In this paper, we report the unequivocal detection of all three optical
atomic oxygen lines in spectra of comet C/1999\,S4 (LINEAR).
Our observations of these three lines suffer from virtually no contamination
from any cometary or telluric line.  We use these data to 
study the ratio of the various transitions and to determine the
widths of the cometary lines.

\section{Observations and Reductions}
Comet C/1999\,S4 (LINEAR) was observed on 10 nights
from 25 June through 17 July 2000 using the 2DCoud\'{e} spectrograph
(Tull {\it et al.} 1995) 
on the 2.7-m Harlan J. Smith telescope at McDonald Observatory.
This cross-dispersed echelle spectrograph, situated at the f/33
coud\'{e} focus of the telescope, was utilized at
a resolving power R=60,000.  In this mode, spectral coverage is complete
from $\sim3700 - 5700$\AA\ with continued coverage with increasing 
interorder gaps to 1\,$\mu$m. 
The detector is a Tektronix 2048x2048 pixel CCD with 24 $\mu$m pixels.
Table~\ref{obslog} lists details of
the observations.  In all cases, the slit projected to $1.2\times8.2$\,arcsec
on the sky and we summed the data along the slit.  At the coud\'{e} focus,
the position angle of the slit on the sky is variable with declination and
hour angle.  During 3 hours of observations it will rotate 45$^\circ$.

In addition to the cometary spectra, observations were obtained
of a ThAr hollow-cathode lamp for calibration of the wavelengths. 
By fitting thorium line positions from all orders, we achieved a dispersion
solution with rms errors of $\sim2.5$\,m\AA.
An incandescent lamp was observed for flat fielding.  
The solar spectrum was observed with an identical instrumental setup
to that used for the comets by imaging the Sun through a diffuser
on the roof of the spectrograph slit room and projecting this image through
the slit in the same manner as objects are observed through the telescope.
Thus, we were able to use an {\itshape observed} solar spectrum in our
reductions.  At least once per observing run, we also observed
$\alpha$~Lyr for relative flux calibration of the orders.
Details of our normal reduction procedure can be found in Cochran {\it et al.}
(2000). 

Figure~\ref{green} shows a representative spectrum of comet LINEAR in the
region of the green oxygen line.  This spectrum was obtained on 17 July 2000.
In the upper panel, the spectrum is shown allowing for the full
strength of the telluric oxygen line.  The cometary and telluric oxygen
lines are separated by almost 0.8\AA\ and are clearly resolved. 
In the bottom panel, the Y axis is expanded by a 
factor of 8 to show better the other, weaker features in the spectrum.
The C$_{2}$ (1,2) bandhead is marked.  The cometary O ($^1$S) line is
considerably stronger than the C$_{2}$ bandhead, unlike the situation
with the spectra of other comets shown in figure~\ref{examples}.
Indeed, it is reasonable to assume there is {\it virtually no} contamination
of the cometary oxygen green line by weak C$_{2}$ lines.  Also, it is apparent
that the C$_{2}$ band in comet LINEAR is much less well developed
than for the other comets.  Indeed, we find the spectrum of this
comet to be depleted in C$_{2}$ and C$_{3}$ relative to CN.
This was confirmed by Farnham {\it et al.} (2001).

Figure~\ref{red} shows representative spectra of comet LINEAR for the
regions of the red oxygen doublet lines, again from 17 July 2000.
Clearly, we have
resolved both of these cometary lines from the telluric lines.
All of the other lines in these spectra are attributable to other
species, generally NH$_{2}$.  In the case of the red doublet, the
cometary line is substantially stronger than the telluric line.

In order to understand the nature of the atomic oxygen in this
comet's coma, we measured the intensity ratio of the green line to the
red lines.  To compute this ratio, we first had to remove the
underlying solar continuum spectrum from each observed spectrum.
In addition, the telluric spectrum has O$_2$ absorption features in
the spectral orders of both lines of the red doublet.  

First, we needed to remove the telluric features from the
two red spectra.  We used the
observations of $\alpha$~Lyr, a mostly featureless spectrum, to
define the telluric spectrum.  This spectrum was shifted to the
rest frame of the observed solar spectrum, matched in telluric line depth
and the solar spectrum was divided by
the telluric spectrum to provide a ``pure'' solar spectrum
with no telluric features.  The telluric features were similarly
removed from the cometary spectra, leaving a cometary emission
plus reflected solar continuum spectrum.

Then, the cometary telluric-corrected red spectra and the green spectrum
(which has no telluric absorptions present) were corrected for
the solar continuum.  The observed green solar spectrum and the
telluric-corrected red solar spectra were shifted to the cometary
rest frame, the continuum of the solar spectrum was matched
to the cometary continuum and the solar spectrum was subtracted.

Finally, the resultant spectra were Doppler-shifted to the 
laboratory rest frame.  During all of these processes, we were
careful to preserve the relative flux levels of the different
spectra.  At this point, we had spectra similar to those
shown in Figures~\ref{green} and \ref{red} for each of the
observations listed in Table~\ref{obslog}.

The intensities and FWHM of all the cometary and telluric atomic 
oxygen lines were computed by fitting the observed lines with
Gaussian profiles.  
We tested Lorentzian and Voigt profiles but found that a Gaussian profile
was more suitable.

It is not sufficient to compare the observed counts in the green
line with those in the red line since the blaze function of the
instrument is not perfectly flat.  However, coud\'{e} spectra
cannot be calibrated in a spectrophotometric manner since the
entrance aperture is so small and it is impossible to ensure
all of a standard star's flux enters the instrument.
However, since all of the spectral orders are observed simultaneously,
we can use observations of a flux standard to determine the
{\it relative sensitivity} of two orders. 
This is true as long as the focal plane is reasonably flat.  We
confirmed this by measuring the stellar spatial FWHM of various orders and
found a variation of $<2\%$ in the FWHM across the wavelength range of 
interest.  To compute the
sensitivity correction, we utilized our observations of 
$\alpha$~Lyr.  We measured the mean continuum counts around
the wavelengths of the three oxygen lines of interest.
The ratios of these counts were then compared with the known
flux of $\alpha$~Lyr (T\"{u}g {\it et al.} 1977).
The difference between the observed ratios and the real flux ratios
represent the needed correction factor.  We found that we
did not need to correct the 6300/6364 ratio from the measured
value, but the measured 5577/6300 ratio needed to be increased
by 30\%.  This correction factor was applied to all
of the flux ratios which we discuss in the next section.

\section{Results}

The intensity ($I$) of an emission line is dependent on the
dissociative lifetime of the parent ($\tau_p$), the yield of photodissociation
($\alpha$), the branching ratio for the transition ($\beta$) and
the column density of the parent ($N)$ in the following manner:
\begin{equation}
I = 10^{-6}\,\tau_p^{-1}\,\alpha\,\beta\,N
\end{equation}
We observed and measured both lines of the red doublet of oxygen.
Since these are both transitions from the (2p$^4$) $^1$D state to
the (2p$^4$) $^3$P ground state, the dissociative lifetime of the parent,
the column density of the parent and the
yield of photodissociation should be the same for these two transitions,
so the ratio of the two line strengths should
be the same as the ratio of the branching ratios.

Table~2 of Festou and Feldman (1981) gives branching
ratios for the O ($^1$S) and O ($^1$D) metastable states of
oxygen.  From this table, we can see that the two red lines should
be produced with a ratio of $\sim$3.15 (6300/6364\AA\ line).
Figure~\ref{redrat} shows the ratios which we measured for the red
doublet from our comet observations.  The dotted line marks the
theoretical ratio calculated from the branching ratios.  As can be seen from
inspection of this figure, most of our observations are consistent with the
branching ratio.  The error bars shown are the 1-$\sigma$ photon statistic
errors.  However, the data of 14 July 2001 (JD=2451739) are in marked
disagreement
with the theoretical prediction.  Inspection of these data showed that
a telluric O$_2$ absorption line was exactly coincident with the cometary 
O~($^1$D -- $^3$P) 6300\AA\ emission line on this date due to the geocentric
radial velocity of the comet.  This strong line was impossible to remove
accurately so the measured flux for this line was underestimated and
the ratio of the two red doublet lines is not the theoretical ratio.
Because of this problem with the 6300\AA\ line, for subsequent analaysis,
we adopted a value for the 6300\AA\ line for 14 July which was $3.15\times$ the
flux of the 6364\AA\ line.  In other words, we assumed the theoretical ratio.
Excluding the data from 14 July, we derive a mean value for the red doublet
ratio of $3.03\pm0.14$ (1$\sigma$ error), in good agreement with the somewhat
uncertain branching ratio. 

In comparing the green line to the red doublet, we are no longer 
observing transitions from the same upper state, so the situation
is not as simple as the intensity ratio of the red doublet.
Figure~\ref{greentored} shows the ratio of the intensity of the green
line to the sum of the intensities of the two lines of the red
doublet (with the ``corrected'' value of the 6300\AA\ line flux for
14 July as mentioned above).  
The error bars are just the 1-$\sigma$ photon counting uncertainties
propagated through the calculation.
In general, the ratio is approximately constant.  However, two
data points appear much higher.  These are the data from 50 and 100 arcsec
tailward on 14 July (JD=2451739).  The discrepancy for these two ratios is
due to the substantial error associated with the very small flux in the
emission lines.  Neglecting these two data points, we find
the ratio of the green line to the red doublet intensity is $0.06\pm0.01$.
This assumes that there is no collisional quenching to alter this
ratio.

Our derived green line-to-red doublet lines ratio is not the first such
measurement, but it is the first where the O~($^1$S) line is relatively
uncontaminated
by C$_{2}$.  Previous observations have found ratios of 0.22--0.34
for comet IRAS-Araki-Alcock (W. Cochran 1984) and 0.12--0.15
for comet C/1996\,B2 Hyakutake (Morrison {\it et al.} 1997).
Smith and Schempp observed the O~($^1$S) and the 6300\AA\ O~($^1$D) lines
in comet Halley and derive a ratio of these two lines (without
trying to correct for the contribution of the 6364\AA\ line) of
0.05--0.1.  Thus, except for the lowest of these limits, our derived
line ratio is the lowest value of any comet.  While it is possible that
the line ratio is different for different comets, we note that all
other cometary observations of the O~($^1$S) line have suffered from
some significant C$_{2}$ contamination.

The ratio of the O~($^1$S) intensity to the sum of the O~($^1$D) intensities
can be expressed in a similar form to equation~1.
\begin{equation}
\frac{I_{5577}}{(I_{6300}+I_{6364})} = \frac{\tau_{p-green}^{-1}\,\alpha_{green}\,N_{green}\,\beta_{5577}}{\tau_{p-red}^{-1}\,\alpha_{red}\,N_{red}\,(\beta_{6300}+\beta_{6364})}
\end{equation}
Recall that oxygen can be produced by the photodissociation of any number
of parent molecules, either as a daughter species or a granddaughter
species.  Thus, evaluation of equation~2 requires knowledge of the
parent(s) of the oxygen.  Conversely, knowledge of the intensity
ratio can lend clues to the parent species which give rise
to these transitions.

Festou and Feldman (1981) argued convincingly that the parent of cometary O
must be H$_{2}$O, CO, or CO$_2$ since any other more complex
oxygen-bearing ice (e.g.
HCOOH or H$_2$CO) does not produce oxygen as a first product
of photodissociation but only as part of subsequent decay of radicals.
The production of oxygen by these other ices would produce oxygen
with a radial extent in the coma which is inconsistent with observations.

The nature of the parent(s) of these three lines is not just of academic
interest.  Direct observations of H$_{2}$O in cometary comae
through the Earth's atmosphere are difficult, although for some
very bright comets
water has been detected with the KAO (e. g. Mumma {\it et al.} 1986; Larson
{\it et al.} 1991) or via the ``hot'' bands (Mumma {\it et al.} 1996; Dello
Russo {\it et al.} 2000).
We can use the primary daughter, OH, as a proxy for studying water.
However, observations of
OH are difficult because the (0,0) band of OH is at 3080\AA\ where 
atmospheric transmission is low.
Observations of OH are best done with a UV telescope such as IUE or HST
(e.g. Weaver {\it et al.} 1981; Feldman {\it et al.} 1987), 
but OH can be detected with some ground-based instruments
(e.g. Cochran and Schleicher 1993; A'Hearn {\it et al.} 1995).
Radio observations can also be used to study OH and are excellent
for velocity resolution but generally are lacking in
spatial resolution because of the very long wavelength
and, hence, large beam size (e.g. Bockel\'ee-Morvan 
{\it et al.} 1990; Crovisier 1989; G\'{e}rard 1990).
Sometimes, however, observations of O~($^1$D) are used as a tracer of water
and the column densities of oxygen are converted to H$_{2}$O production
rates under the assumption that all of the oxygen comes from H$_{2}$O
(e.g. Spinrad, 1982; Fink and DiSanti, 1990; Magee-Sauer {\it et al.} 1990;
Schultz {\it et al.} 1992).  

Our observations of both the green oxygen line and
the red doublet can verify the validity of the assumption that all of the 
($^1$D) state comes from photodissociation of H$_{2}$O.
If we assume that there is only \underline{one} dominant parent of oxygen, then
the column densities of the parent in the numerator and denominator of
equation~2 are identical.  Then, the effective excitation rate for
photodissociation of a parent molecule is proportional to $\tau^{-1}\,\alpha\,\beta$.
These excitation rates for H$_{2}$O, CO, and CO$_2$ are given in
Table~\ref{excite}.
Examination of the effective excitation rates in Table~\ref{excite} in
comparison with our derived ratio of 0.06 indicates that the likely
parent of the forbidden oxygen lines in the coma of comet LINEAR is
H$_{2}$O.
For CO$_2$, Festou and Feldman (1981) and Delsemme (1980) derive 
very different values from one another for the green to red line ratio. 
However, neither
value is consistent with our observed value of 0.06.

In addition to measuring the ratio of the line intensities, we measured
the widths of the cometary O lines.  An observed line has a width which is the 
convolution of the instrumental line width and the velocity line width
in the coma.  In order to determine the actual cometary
line width, therefore, the measured line width must be corrected
for the instrumental line widths.

We measured the instrumental line widths for each echelle order by using
our observations of the ThAr hollow-cathode lamp.
The intrinsic widths of the Th lines are very narrow.  The lines are
only just resolved with the coud\'{e} spectrograph at R=500,000, so they
can be used to measure accurately the instrumental profile at R=60,000.
The hollow-cathode lamp optics were designed to give a pupil matching
that of the telescope and to illuminate the slit similarly to an
object.  Thus, the observed Th line widths are excellent measures of the
spectrograph instrumental profile width.
The instrumental widths
are a function of wavelength, so for each spectral order we measured
many thorium line widths.  This procedure was completed for at least
one arc lamp spectrum per night.  Next, we measured the line widths
for all three oxygen lines in each cometary spectrum.  From these
measurements, we could compute the average values for the instrumental
width and the convolved widths and could determine a deconvolved
oxygen width for each line by subtracting the instrumental from the
measured line widths in quadrature. 

Table~\ref{widths} lists our
results.  Examination of this table shows that all three oxygen
lines are wider than the instrumental resolution.  This added
line width is a measure of the velocity dispersion of the gas in the coma.
Our width for the 6300\AA\ line is consistent with the width of the same
line in Kohoutek (1973\,XII; Huppler {\it et al.} 1975)
though our error bars are much smaller.
Smyth {\it et al.} (1995) used observations with R=190,000 and found
OI 6300\AA\ line widths of 2.07-3.22 km/sec, also in good agreement.
Their data were obtained at heliocentric distances of 0.8 and 1.6\,{\sc au}.
 
\section{Discussion}

We have measured the intensity ratio of the ($^1$S) to ($^1$D) lines.
We can use the effective excitation
rates of Table~\ref{excite} to infer the parent. 
However, inspection of this table shows
that these rates are not constrained well.  Indeed, with the rates
quoted, it can be seen that CO$_2$ is a more efficient producer
of O~($^1$S) than H$_{2}$O.  Therefore, we investigated
other details about this comet and about the rates which might have
an affect on our conclusions.

A potential parent for the oxygen is the photodissociation of CO.
Evidence of CO in cometary comae can be found with the CO Fourth Positive
group in the UV, with the IR CO bands and with CO$^+$ in the tail of comets.
All three of these lines of evidence were examined for this comet
and it was shown that comet LINEAR was depleted in CO relative
to other comets.  We used our tailwards observations 
to search for the CO$^+$.  We detected no CO$^+$ in any of our
spectra.  However, a non-detection of this ion is not a strong
constraint on the quantity of CO since ion tails can be quite
narrow or even absent and therefore we may not have sampled the ion tail.

HST observations of LINEAR using the Space Telescope
Imaging Spectrograph (STIS) on 5 July 2000 were used to detect several
lines of the CO Fourth Positive group and to derive a production
rate of CO of $\sim5\times10^{26}$\,sec$^{-1}$ (Weaver {\it et al.} 2001). 
This yields an {\itshape upper limit} of CO/H$_{2}$O of 0.6\%. 
On that same date, CSHELL was used on the IRTF to detect the 
R1 line of the (1,0) band of CO at $\sim2151$\,cm$^{-1}$ and to derive
a CO production rate  of $7\pm2\times10^{26}$\,sec$^{-1}$
(Mumma {\it et al.} 2001a).
The derived yield of CO/H$_{2}$O is 0.9\%.
These derived yields for comet LINEAR can be compared with comets
Lee (1.8\%; Mumma {\it et al.} 2001b), Halley (3.5\%; Eberhardt 1999),
Hyakutake ($\sim$10--14\%, Mumma {\it et al.} 2001a) and Hale-Bopp 
(12.3\%; DiSanti {\it et al.} 1999, 
2001) to show that LINEAR is depleted in CO relative
to H$_{2}$O when compared with other comets.
Indeed, in comet LINEAR, the yield of CO is so low that, coupled
with the low excitation rates, it would be difficult
to produce much oxygen from CO relative to that which is produced from
H$_{2}$O.  Thus, we conclude from the low yield and the fact that for a CO
parent the green and red lines should be in a ratio of 1:1 that CO is not
an important parent of the oxygen in comet LINEAR.

Direct observational evidence for the presence of CO$_2$ in cometary comae
is more difficult than for CO.
CO$_2$ is a linear symmetric molecule and its ground state does not
absorb photons in the visible or UV and it has no allowed radio
transitions.  The IR transitions
cannot be detected except from above the Earth's atmosphere. 
Some high-vibrational overtones of the IR fundamental bands will fall
in the near-IR windows, but these overtone bands are very weak.
Thus, there are no observed CO$_2$
bands in cometary spectra.  However, the presence of CO$_2$ can be
deduced in two ways: CO$_2^+$ emissions in the UV and visible spectrum
of the tail, and the Cameron bands of CO in the UV (1990--2160\AA). 
Generally, the CO$_2^+$ bands are difficult to detect because they are
weak and occur in the near-UV.  They are not included in our
bandpass so we cannot comment on if they were present.

The Cameron CO bands arise from photodissociation of CO$_2$ (Lawrence 1972;
Weaver {\it et al.} 1994)
and therefore represent a good proxy for the CO$_2$. 
HST with STIS was used to search for these bands but the 
Cameron bands were not detected (Weaver, personal communication, 2001).
HST is $10\times$ less sensitive at the Cameron band wavelengths
than at the wavelengths of the CO Fourth Positive band, so the failure
to detect the Cameron bands does not definitively mean that there
are no Cameron band emissions.  Still, a preliminary
estimate by Weaver is that CO$_2$/H$_{2}$O is $\leq5\%$ for
comet LINEAR. 
This ratio is slightly lower than or comparable to that found for
five other comets using data from the IUE
(Feldman {\it et al.} 1997) and HST (Weaver {\it et al.} (1994).
Feldman {\it et al.} also found that CO$_2$/CO ratios of $>$1 are common.
Inspection of Table~\ref{excite} (line 3) shows that
CO$_2$ is a more efficient producer of O ($^1$S) than H$_{2}$O, so even
with the low CO$_2$/H$_{2}$O implied by the STIS observations, CO$_2$
could be a significant source of O ($^1$S).

We have ignored the effects of solar activity in this study, even though
the Sun was quite active during the time of our observations.
The photodissociation of most species is sensitive to the solar
ultraviolet flux.  Flux in discrete UV wavelength bins are responsible
for the photodissociation of different branches.  A rate coefficient, k,
for a wavelength interval from $\lambda_1$ to $\lambda_2$ can be
computed from
\begin{equation}
\mathrm{k} (\Delta\lambda) = \int^{\lambda_2}_{\lambda_1} \sigma(\lambda) \Phi(\lambda) d\lambda 
\end{equation}
where $\sigma(\lambda)$ is a photo absorbtion cross-section and $\Phi(\lambda)$
is the photon flux at wavelength $\lambda$.

Festou and Feldman's (1981) excitation rates for CO$_2$
utilized the quiet Sun flux of Huebner and Carpenter (1979) and the
photodissociation cross-sections of Lawrence (1972a,b).
The cross-sections are for wavelengths bluer than Ly~$\alpha$.
For Delsemme's (1980) line ratio, the solar flux and cross-sections are
illustrated in his Figure~2.
The caption indicates the solar flux is for the ``mean'' Sun.
Oppenheimer and Downey (1980) pointed out that the solar UV flux is
quite variable during the solar cycle and can cause a change in
the excitation rates of a factor of two or more.
Budzien {\it et al.} (1994) showed that near solar maximum it is
important to consider the short-term variability of the Sun since the
amplitude of the 27-day variation in some solar parameters approaches
that of the 11-year cycle of activity (see their Figure~3).

Since CO$_2$ is such an important constituent for the atmospheres
of Mars and Venus, several newer analyses have been done
of the photo absorption cross-sections (though the conditions in
these atmospheres are very different than in the cometary coma so
that derived photodissociation rate coefficients from these
studies are not applicable to comets).
Lewis and Carver (1983) have derived new absorption cross-sections
from 1200 to 1970\AA\ including measuring the effects of temperature
on the cross-sections (measurements were made at 200, 300, and 370K).  
They found that the temperature effect is small at the shorter
wavelengths, passing through a minimum at 1400\AA, but that the cross-section
can vary by as much as a factor of 20 at 1900\AA\ when raising the
temperature from 200 to 370K.
In addition, Anbar {\it et al.} (1993) pointed out that the cross-section
measurements show large variations in 10-20\AA\ scales so that calculations
of the excitation rates can be sensitive to the wavelength resolution.
The extent to which these two factors (resolution and temperature) are
important to comets is presently unknown. Also unknown is how important
the wavelength channel of the new cross-sections,
relative to the FUV channel measured by Lawrence (1972a,b),
are for the photodissociations needed to produce the oxygen. 
Detailed calculations will be necessary to quantify this effect, though the
calculation of new excitation factors is outside the scope of this paper.

For oxygen produced from the photolysis of H$_{2}$O, the range
of effective excitation rates listed in Table~\ref{excite} probably
encompasses a range of solar activity.  Festou (1981) showed
that the relative importance of the different photodissociation channels varies
with solar activity.  Lyman $\alpha$ increases from 24\% of
the pathway for a quiet Sun to 39\% for an active Sun, while
the photodissociation branch from 1357--1860\AA\ decreases from
72\% to 58\% with the increase in activity.  The channel with
$\lambda<1357$\AA\ (but not including Ly $\alpha$) stays roughly
constant at 3--4\%.  Within these channels, the branches that
produce the oxygen lines also change relative importance so that
in total, the ($^1$D) is produced 6.7\% of the time for a quiet
Sun and 9.8\% of the time for an active Sun, while the numbers
for ($^1$S) are 0.6\% and 1.0\%.  Fortuitously, the
relative ratio of these two lines is about the same for either
a quiet or active Sun.  Crovisier (1989) agreed on the relative
branching to produce oxygen for the two non-Ly $\alpha$ channels
but found that Ly $\alpha$ dissociated only 10\% of the time
to oxygen while Festou lists this branch as active 25\% of the time.
Budzien {\it et al.} (1994) found
that the photodestruction rates can vary by 30\% from solar minimum
to maximum, but concluded that the quantum yield of OH and O ($^1$D)
from H$_{2}$O photodissociation is relatively insensitive to solar
flux variations.  Since we are trying to determine the relative
production of the oxygen green and red lines, it is probably reasonable
to ignore the details of the solar activity for our study.

Thus, our data show a ratio of green to red intensity of 0.06 and
this value is relatively constant despite the fact that the 
comet was quite variable and was, in fact, disintegrating.
The constancy of this ratio, despite the activity, indicates that the scale
lengths of the parent of each state are very similar.  This, however,
does not preclude separate parents for the two states.

The line intensity ratio of 0.06 is inconsistent with CO$_2$ as the dominant
parent using the effective excitation rates of CO$_2$ of either Festou and
Feldman (1981) or Delsemme (1980), although the predicted ratios
in these two works are quite different.  The intensity ratio predicted
for H$_{2}$O is consistent with the observed ratio since the
effective excitation rates given span a range.  However, the caveats
mentioned above about solar flux and cross-sections must be
remembered when drawing any conclusions.

Regardless of the actual parent which gives rise to the oxygen transitions
which we have detected, the determined line widths indicate something
about the nature of the gas which populates these states.
Recall that the oxygen atom can be formed in either the ground state
or the excited state.  95\% of the atoms which are in the
($^1$S) state will decay back to the ground state via the
($^1$D) state and while doing so they will produce ($^1$D) lines
which are {\itshape as wide as} the ($^1$S) line.

We measured the ($^1$S)/($^1$D) flux ratio to be 0.06.  Thus, most of the
atoms formed in the ($^1$D) state.  However, the ($^1$S) line is
wider than the ($^1$D) lines and both lines of the red doublet are
of a consistent width.  These two facts, in combination, imply that
atoms which start in the ($^1$D) state form lines with a {\it lower
velocity dispersion} (narrower) than atoms which start in the ($^1$S) state.

However, recall also that the lifetime 
of the $^1$D state at 1\,{\sc au} is about 150\,sec, while for the $^1$S state 
it is less than 1 sec.  Both are extremely short and, as a result, the
line widths are a measure of the velocity dispersion of the parent, not
the oxygen;
the wider green line implies that its parent has a
higher velocity dispersion than the red doublet parent.
This is suggestive that the parent of the O in the ($^1$S) state and the parent
of the majority of the O in the ($^1$D) state are \underline{not} the same.

\section{Summary}
In this paper, we have reported on high spectral resolution observations
of comet C/1994\,S4 (LINEAR) with which we have detected unequivocally
three oxygen metastable lines.  
\begin{enumerate}
\item We have derived the ratio of the line strengths of the two lines of
the red oxygen doublet and have shown that this ratio is as 
predicted by the branching ratios commonly reported in the literature.

\item We have derived the ratio of the line strength of the green line
to the sum of the strengths of the red lines and shown that it was
approximately constant with a value of $0.06\pm0.01$.  This
ratio is consistent with H$_{2}$O as the dominant parent for the atomic oxygen.
However, based on line width measurments,
there may be another parent contributing to the oxygen.

\item We have measured the widths of the three oxygen lines and have
shown that the green line is wider than the red lines, implying a
higher velocity dispersion for the upper level than the lower level.
Since 95\% of the upper level (green line) atoms decay to the ground state via
the red doublet, this implies that atoms which originate in the upper
state have a higher velocity dispersion than those which originate
in the lower excited state.

\item We discussed the implications of solar activity and of the various
physical parameters for our conclusions.
Solar activity is probably a secondary effect that will not change our
conclusions.  New rate coefficients for the photodissociation of CO$_2$ to 
oxygen will need to be calculated before CO$_2$ can be eliminated as a 
contributing parent on the basis of the line ratios.

\end{enumerate}

\vspace{10pt}
\begin{center}Acknowledgements\end{center}
This work was funded by NASA Grant NAG5 9003.
We thank Michael Combi and Michael Mumma for helpful discussions.

\clearpage
\begin{table}
\caption[obslog]{Observing Parameters}\label{obslog}
\centering
\begin{tabular}{r@{ }l@{ }lr@{:}l@{\ \ }cr@{.}lcccl}    
\hline
 \\ [-10pt]
 & & &\multicolumn{2}{c}{UT}   & R$_h$ & \multicolumn{2}{c}{$\dot{\rm R}_h$}  &$\Delta$ & Exposure & PA & \multicolumn{1}{c}{Position of}\\
\multicolumn{3}{c}{Date} &\multicolumn{2}{c}{Start} & 
({\sc au}) & \multicolumn{2}{c}{(km\,sec$^{-1}$)} & ({\sc au}) & Time (s) & Tail$^1$& \multicolumn{1}{c}{Slit$^2$} \\
\hline
25 & Jun & 2000 & 08 & 47 & 0.97 & \mbox{\hspace*{1em}}-19 & 7 & 1.21 & 1800 & 270.7 & optocenter \\
   &     &      & 09 & 25 &      & \multicolumn{2}{c}{ } &      & 1800 &       & optocenter \\
   &     &      & 09 & 59 &      & \multicolumn{2}{c}{ } &      & 1800 &       & optocenter \\
   &     &      & 10 & 33 &      & \multicolumn{2}{c}{ } &      & 1800 &       & optocenter \\
26 & Jun & 2000 & 08 & 26 & 0.96 & -19 & 4 & 1.17 & 1800 & 270.7 & optocenter \\
   &     &      & 09 & 05 &      & \multicolumn{2}{c}{ } &      & 1800 &       & optocenter \\
   &     &      & 09 & 39 &      & \multicolumn{2}{c}{ } &      & 1800 &       & optocenter \\
   &     &      & 10 & 12 &      & \multicolumn{2}{c}{ } &      & 1800 &       & optocenter \\
   &     &      & 10 & 46 &      & \multicolumn{2}{c}{ } &      & 1200 &       & optocenter \\
06 & Jul & 2000 & 09 & 03 & 0.86 & -15 & 1 & 0.81 & 1800 & 274.6 & optocenter \\
   &     &      & 09 & 38 &      & \multicolumn{2}{c}{ } &      & 1800 &       & optocenter \\
   &     &      & 10 & 13 &      & \multicolumn{2}{c}{ } &      & 1800 &       & optocenter \\
   &     &      & 10 & 49 &      & \multicolumn{2}{c}{ } &      & 1800 &       & optocenter \\
07 & Jul & 2000 & 09 & 07 & 0.85 & -14 & 5 & 0.77 & 1800 & 275.7 & optocenter \\
   &     &      & 09 & 41 &      & \multicolumn{2}{c}{ } &      & 1800 &       & optocenter \\
   &     &      & 10 & 29 &      & \multicolumn{2}{c}{ } &      & 1800 &       & optocenter \\
   &     &      & 10 & 53 &      & \multicolumn{2}{c}{ } &      & 1200 &       & optocenter \\
08 & Jul & 2000 & 08 & 46 & 0.84 & -14 & 0 & 0.74 & 1800 & 277.1 & optocenter \\
   &     &      & 09 & 20 &      & \multicolumn{2}{c}{ } &      & 1800 &       & optocenter \\
09 & Jul & 2000 & 08 & 47 & 0.84 & -13 & 4 & 0.70 & 1800 & 278.7 & optocenter \\
14 & Jul & 2000 & 09 & 04 & 0.80 & -10 & 0 & 0.53 & 1800 & 295.8 & optocenter \\
   &     &      & 09 & 41 &      & \multicolumn{2}{c}{ } &      & 1800 &       & 50 arcsec west \\
   &     &      & 10 & 17 &      & \multicolumn{2}{c}{ } &      & 1800 &       &100 arcsec west \\
   &     &      & 10 & 51 &      & \multicolumn{2}{c}{ } &      & 1200 &       & optocenter \\
15 & Jul & 2000 & 09 & 59 & 0.80 &  -9 & 3 & 0.50 & 1200 & 302.5 & optocenter \\
16 & Jul & 2000 & 10 & 23 & 0.79 & \multicolumn{2}{c}{ } & 0.48 & 1800 & 311.0 & optocenter \\
   &     &      & 10 & 57 &      & \multicolumn{2}{c}{ } &      & 1200 &       & optocenter \\
17 & Jul & 2000 & 08 & 22 & 0.79 &  -7 & 7 & 0.45 & 1800 & 321.8 & optocenter \\
   &     &      & 08 & 59 &      & \multicolumn{2}{c}{ } &      & 1800 &       & 25 arcsec west \\
   &     &      & 09 & 49 &      & \multicolumn{2}{c}{ } &      & 1800 &       & 10 arcsec west, 10 north \\
   &     &      & 10 & 24 &      & \multicolumn{2}{c}{ } &      & 1800 &       & 10 arcsec east, 10 south \\
   &     &      & 10 & 57 &      & \multicolumn{2}{c}{ } &      & 1200 &       & optocenter \\
\hline
\multicolumn{3}{l}{Notes: } \\
 & 1 & \multicolumn{10}{l}{Position angle of predicted extended heliocentric radius vector (north through east)} \\
 & 2 & \multicolumn{10}{l}{Slit position relative to optocenter} \\
\end{tabular}
\end{table}

\begin{table}
\caption{Effective Excitation Rates for Dissociation}\label{excite}
\centering
\begin{tabular}{lllc}
\\
\hline
\multicolumn{1}{c}{Parent} & \multicolumn{2}{c}{Excitation Rate} & Ratio \\
 & \multicolumn{2}{c}{(sec$^{-1}$ at 1\,{\sc au})} \\
\cline{2-3}
 \\ [-9pt]
 & \multicolumn{1}{c}{O $^1$S} & \multicolumn{1}{c}{O $^1$D} & O ($^1$S)/ O ($^1$D) \\
\hline
H$_{2}$O$^*$ & $7 - 12\times10^{-8}$ & $8 - 12\times10^{-7}$ & $\sim$0.1 \\
CO$^*$ & $<4\times10^{-8}$ & $<4\times10^{-8}$ & $\sim$1 \\
CO$_2$$^*$ & $4.4\times10^{-7}$ & $5\times10^{-7}$ & $\sim$1 \\
CO$_2$$^\dagger$ & & & $\sim0.3$ \\
\hline
\multicolumn{4}{l}{$^*$Source: Festou and Feldman (1981), Table~3} \\
\multicolumn{4}{l}{$^\dagger$Source: Delsemme (1980)}
\end{tabular}

\end{table}
\clearpage
\begin{table}
\caption{The Oxygen Line Widths}\label{widths}
\centering
\begin{tabular}{lcccc}
\hline
\multicolumn{1}{c}{Transition}
& Measured & Measured & Intrinsic & Derived \\
 & Cometary & Instrumental & Cometary & Outflow \\
& Widths & Widths & Widths & Velocity \\
& (\AA) & (\AA) & (\AA) & (km sec$^{-1}$) \\
\hline
O ($^1$S) 5577\AA & $0.110\pm0.006$ & $0.087\pm0.003$ & $0.067\pm0.003$ & 3.60$\pm$0.16 \\
O ($^1$D) 6300\AA & $0.107\pm0.005$ & $0.094\pm0.003$ & $0.051\pm0.003$ & 2.43$\pm$0.14 \\
O ($^1$D) 6364\AA & $0.118\pm0.004$ & $0.103\pm0.005$ & $0.058\pm0.003$ & 2.73$\pm$0.14 \\
\hline
\end{tabular}
\end{table}

\clearpage
\section{Figure Captions}

\begin{figure}[p]
\vspace{7in}
\includegraphics{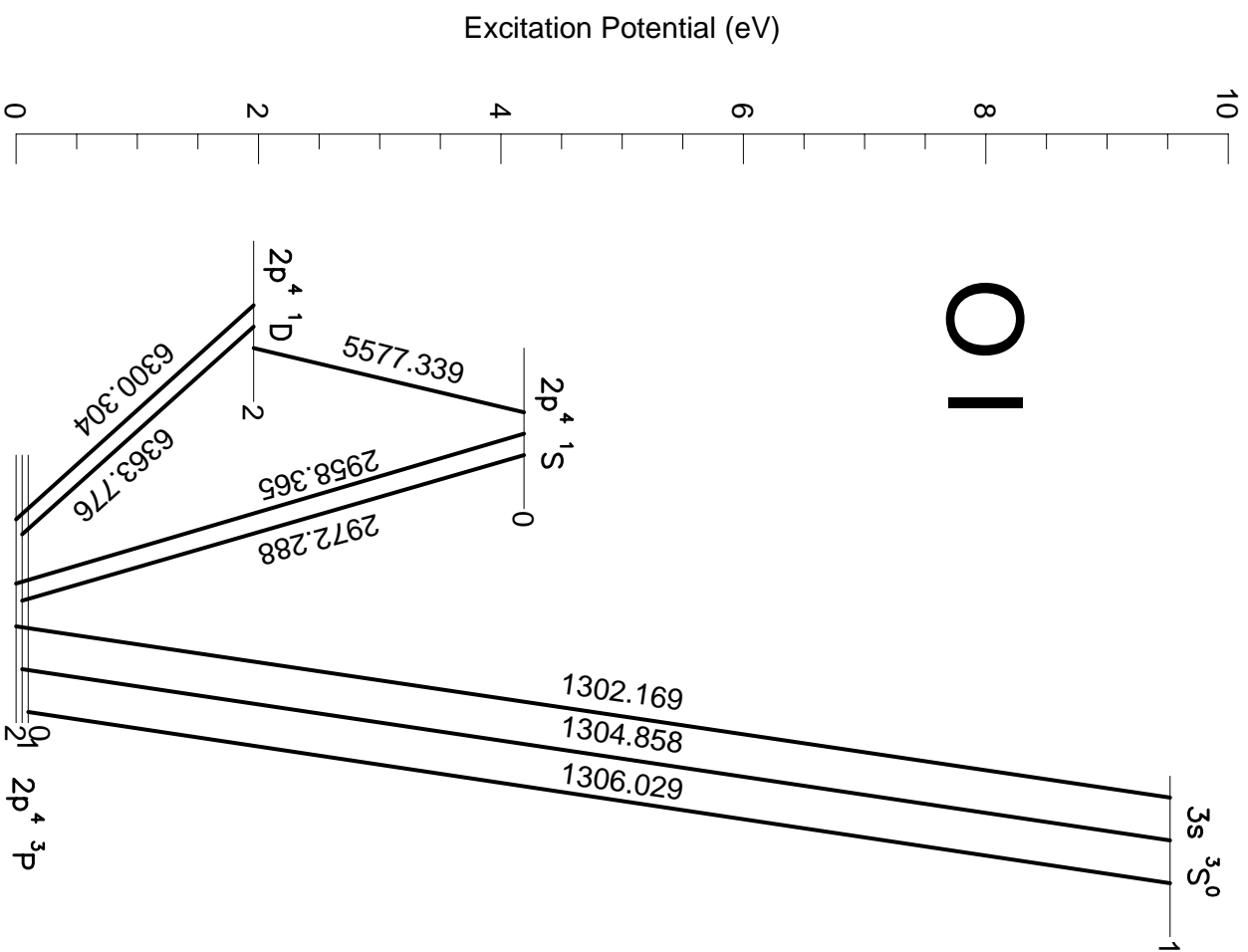}
\caption[fig1]{Cochran and Cochran 2001}\label{transit}
\end{figure}

\noindent
{\bf Figure~\ref{transit}:} An energy level diagram of atomic oxygen
for the lowest
level transitions to the ground state.  The ground state splitting
has been exagerated so the splitting is apparent.  The J=1 level is
really 0.02eV above the J=2 level, while the J=0 level is 0.03eV
above the J=2 level.  The $^1$D--$^3$P J=2-0 transition at
6391.733\AA\ is not shown since it is only predicted and has not
been detected.  The wavelengths are given in \AA ngstroms.

\begin{figure}[p]
\vspace{7in}
\includegraphics{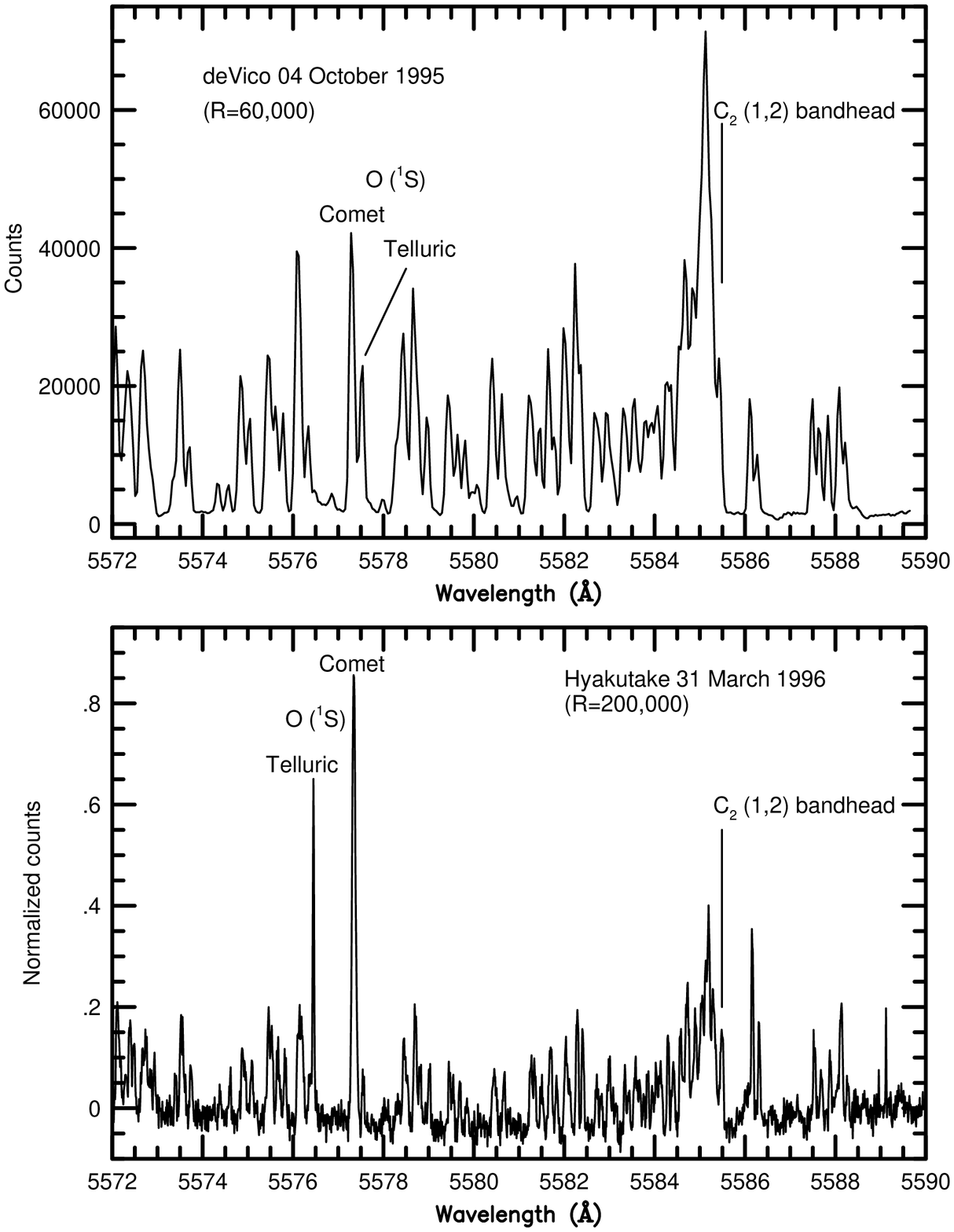}
\caption[fig2]{Cochran and Cochran 2001}\label{examples}
\end{figure}

\noindent
{\bf Figure~\ref{examples}:} High spectral resolution observations of
other comets.  Observations of 122P/1995\,S1 (deVico) are shown in the upper
panel
with R=60,000. 
The slit covered 870km$\times$5845km centered on the optocenter.
The positions of the cometary and telluric lines are shown. 
Obviously, both the cometary and telluric O~($^1$S) are
heavily contaminated by C$_{2}$.  Observations of C/1996\,B2 (Hyakutake)
are shown in the bottom panel.  The spectral resolution is higher than
for deVico; the comet was quite close to the Earth so the spatial
resolution is high.  
The slit covered 46km$\times$1249km centered on the optocenter.
Therefore, there
is less likely to be C$_{2}$ in the aperture and the O ($^1$S) lines
are less contaminated than for deVico.

\begin{figure}[p]
\vspace{7in}
\includegraphics{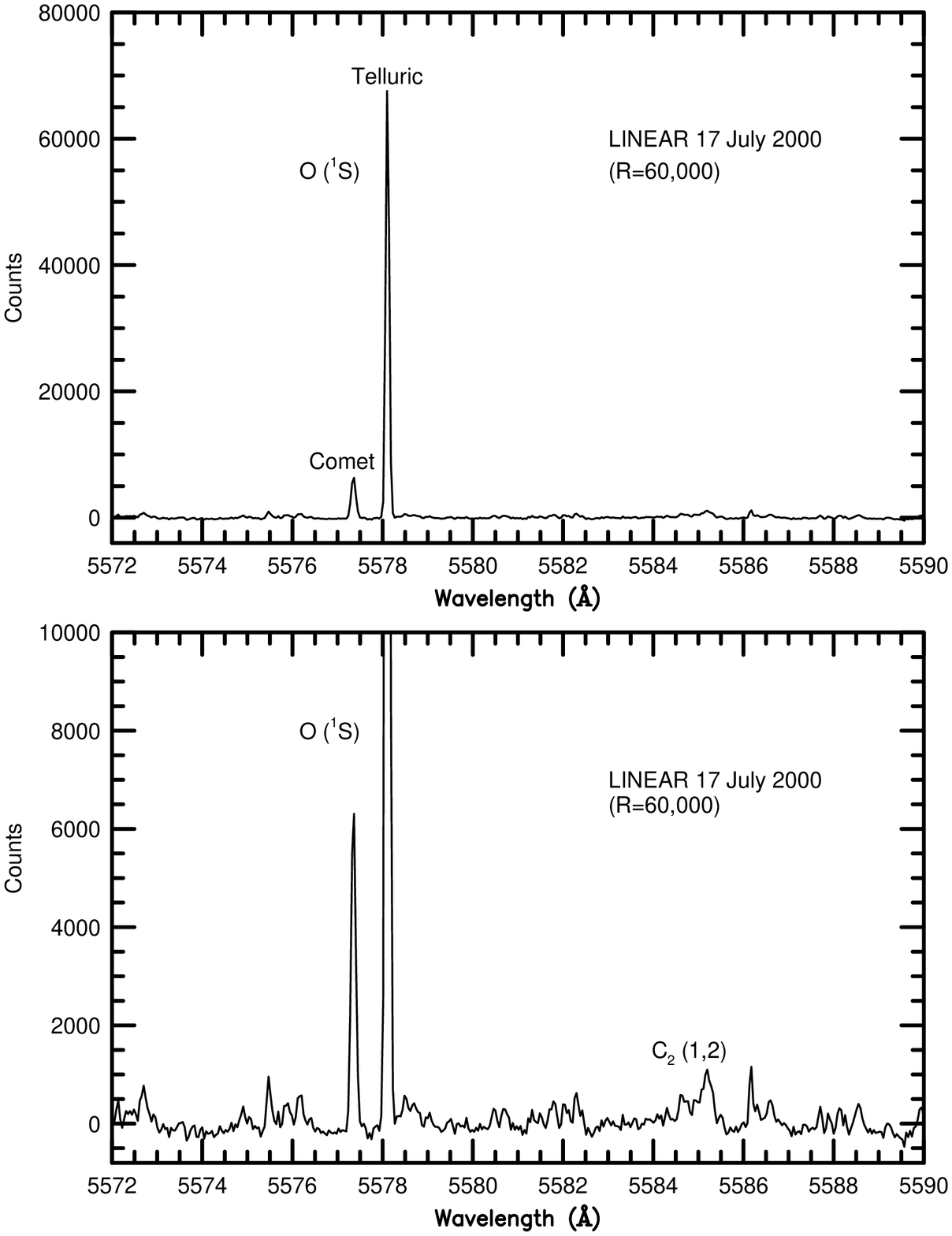}
\caption[fig3]{Cochran and Cochran 2001}\label{green}
\end{figure}

\noindent 
{\bf Figure~\ref{green}:} The region of the O ($^1$S) oxygen line.
This spectrum of the optocenter of LINEAR was obtained on 17 July 2000.
The upper panel shows the spectral region scaled to the telluric
line.  The cometary line is clearly well separated from the telluric one.
The lower panel shows an expansion of the Y axis.  The C$_{2}$ (1,2)
bandhead is marked.  The cometary O ($^1$S) is much stronger even
than the C$_{2}$ bandhead.  It is clear that there is virtually
no contamination of the cometary O ($^1$S) line by cometary C$_{2}$ or
telluric O~($^1$S).

\begin{figure}[p]
\vspace{7in}
\includegraphics{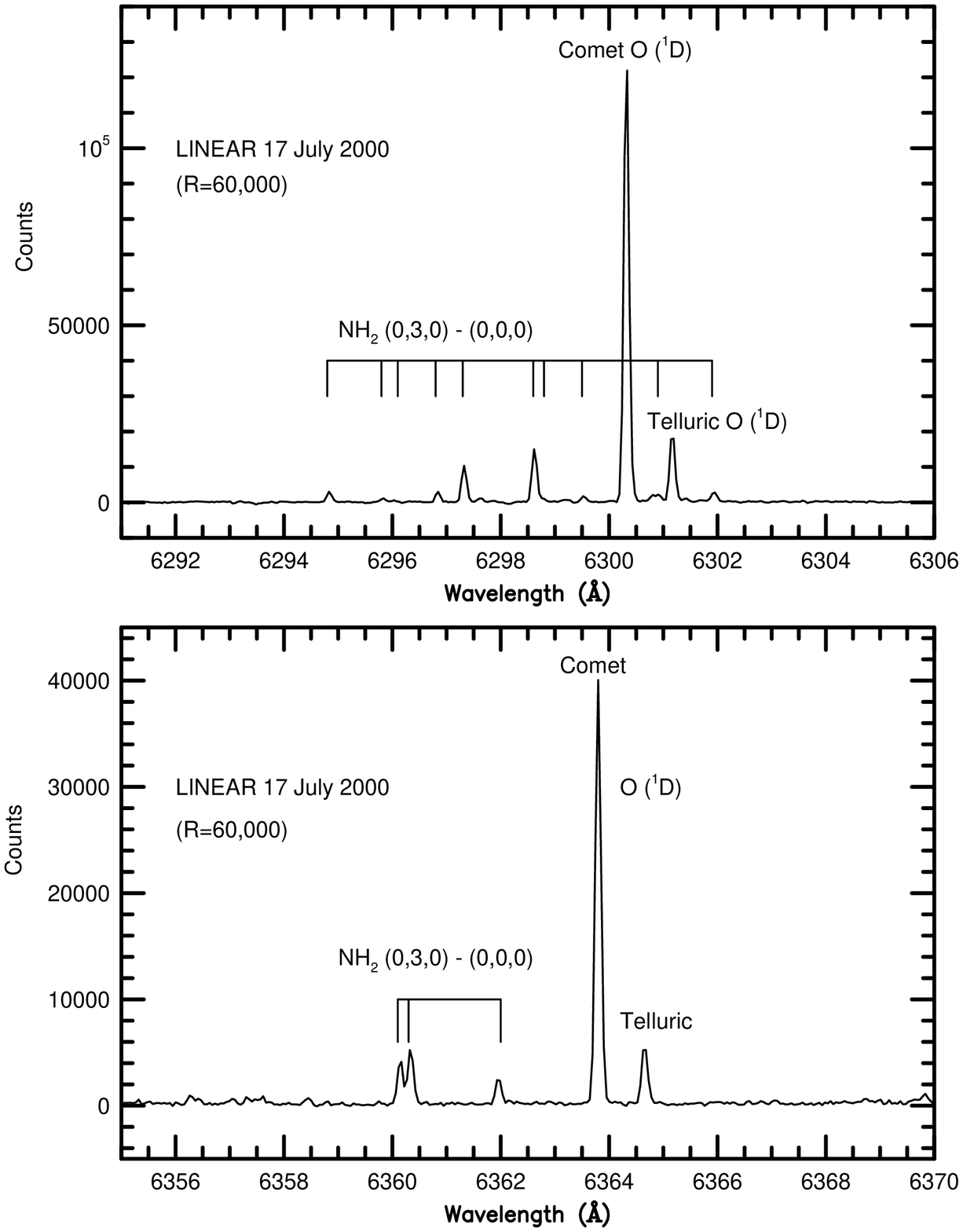}
\caption[fig4]{Cochran and Cochran 2001}\label{red}
\end{figure}

\noindent
{\bf Figure~\ref{red}:} The regions of the O ($^1$D) oxygen lines.
The 6300\AA\ line region is shown in the upper panel, while the 
6364\AA\ line region is shown in the lower.  Again, the cometary
and telluric lines are well separated and no other species contaminates
the cometary lines.  The NH$_{2}$ band designations are in the bent
notation.

\begin{figure}[p]
\vspace{7in}
\includegraphics{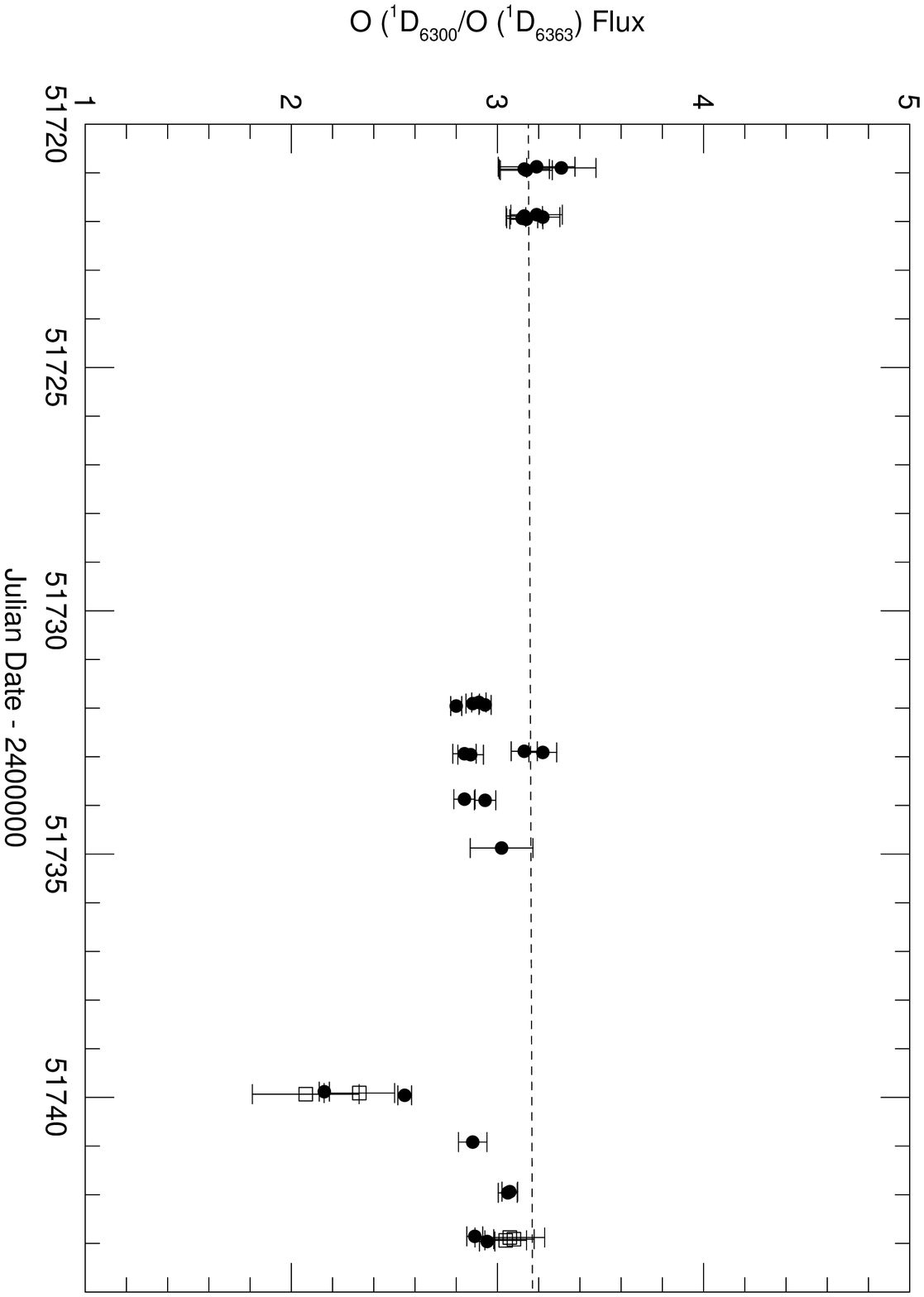}
\caption[fig5]{Cochran and Cochran 2001}\label{redrat}
\end{figure}

\noindent
{\bf Figure~\ref{redrat}:} The ratio of the observed 6300\AA\ red line
to the 6364\AA\ red line.  The cometary optocenter ratios are shown as
solid dots and the non-optocenter ratios are shown as open squares.
The theoretical ratio is denoted by a dashed line.  The data uphold the
theoretical prediction with the exception of the four data points from 14 July
(JD=2451739).
These points are discussed in the text.

\begin{figure}[p]
\vspace{7in}
\includegraphics{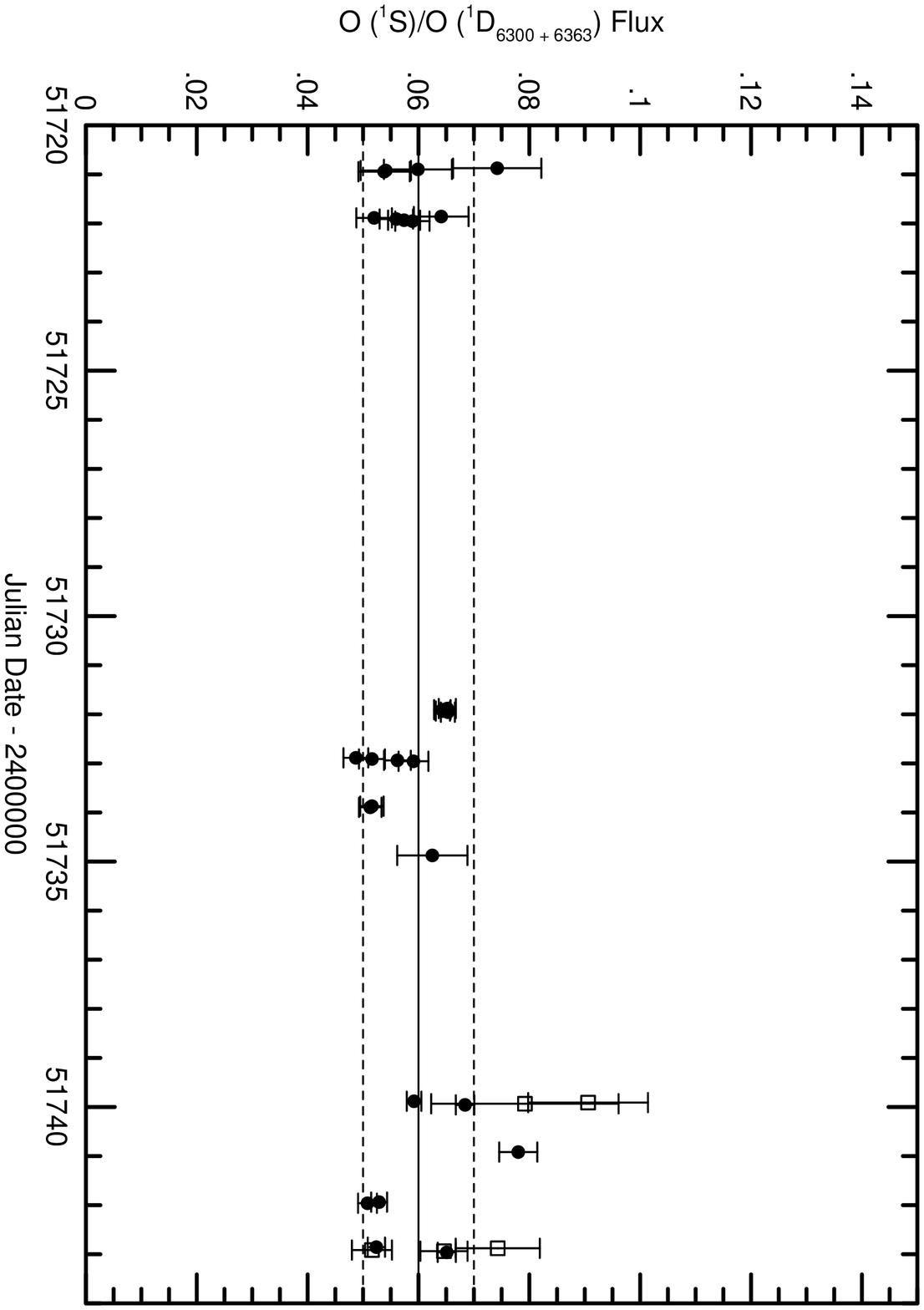}
\caption[fig6]{Cochran and Cochran 2001}\label{greentored}
\end{figure}

\noindent
{\bf Figure~\ref{greentored}:} The ratio of the observed green line to
the sum of the red lines. 
The symbols are the same as for figure~\ref{redrat}.
The two high values of the ratio are the low signal/noise
points at 50 and 100 arcsec tailward from 14 July.  Except for these
two data points, the green-to-red ratio is approximately constant,
with a value of $0.06\pm0.01$. The mean is denoted by a solid line
with the error envelope noted with dashed lines.

\clearpage

\end{document}